\documentclass[sigconf]{acmart}
\acmConference[EASE 2025]{The 29th International Conference on Evaluation and Assessment in Software Engineering}{17–20 June, 2025}{Istanbul, Türkiye}

\usepackage{booktabs}
\usepackage{multirow}
\usepackage{makecell}

\usepackage{xcolor} 

\makeatletter
\def\@copyrightpermission{\footnotesize This is the author's version of the work. It is posted here for your personal use.}
\makeatother

\title{Tracking the Moving Target: A Framework for Continuous Evaluation of LLM Test Generation in Industry}

\begin{document} 
\author{Maider Azanza}
\email{maider.azanza@ehu.eus}
\orcid{0000-0002-4537-1572}
\affiliation{
  \institution{University of the Basque Country UPV/EHU}
  \city{San Sebastian}
  \country{Spain}
}

\author{Beatriz Pérez Lamancha}
\email{bperez@lksnext.com}
\orcid{0009-0001-0672-8081}
\affiliation{
  \institution{LKS Next}
  \city{Mondragón}
  \country{Spain}
}

\author{Eneko Pizarro}
\email{epizarro001@ikasle.ehu.eus}
\orcid{0009-0001-6295-4067}
\affiliation{
  \institution{University of the Basque Country UPV/EHU}
  \city{San Sebastian}
  \country{Spain}
}

\begin{abstract}
\textit{Large Language Models (LLMs)} have shown great potential in automating software testing tasks, including test generation. However, their rapid evolution poses a critical challenge for companies implementing DevSecOps - evaluations of their effectiveness quickly become outdated, making it difficult to assess their reliability for production use. While academic research has extensively studied LLM-based test generation, evaluations typically provide point-in-time analyses using academic benchmarks. Such evaluations do not address the practical needs of companies who must continuously assess tool reliability and integration with existing development practices. This work presents a measurement framework for the continuous evaluation of commercial LLM test generators in industrial environments. We demonstrate its effectiveness through a longitudinal study at \textit{LKS Next}. The framework integrates with industry-standard tools like SonarQube and provides metrics that evaluate both technical adequacy (e.g., test coverage) and practical considerations (e.g., maintainability or expert assessment). Our methodology incorporates strategies for test case selection, prompt engineering, and measurement infrastructure, addressing challenges such as data leakage and reproducibility. Results highlight both the rapid evolution of LLM capabilities and critical factors for successful industrial adoption, offering practical guidance for companies seeking to integrate these technologies into their development pipelines.
\end{abstract}



\received{20 February 2007}
\received[revised]{12 March 2009}
\received[accepted]{5 June 2009}

\maketitle

\section{Introduction}

Software testing remains a critical yet resource-intensive aspect of modern software development. Recent surveys reveal that writing tests is one of the tasks developers enjoy least and would most like to delegate to artificial intelligence \cite{Sergeyuk2025}. This reluctance to write tests, combined with their essential role in software quality assurance, creates a tension in software development practices - particularly in DevSecOps environments where rapid iteration and continuous delivery demand both speed and reliability.

With the emergence of \textit{Large Language Models (LLMs)}, companies are exploring their potential to automate test generation and reduce the testing burden on developers. While recent research has extensively studied LLM-based test generation \cite{Wang24}, demonstrating promising results in areas such as unit test or test oracle generation, industrial adoption faces challenges that extend beyond those addressed in academic studies.

A critical aspect companies must address when adopting LLM-based testing tools lies in their rapid evolution - what was true about a tool's capabilities six months ago may no longer reflect its current performance. This is particularly challenging for organizations implementing \textit{Continuous Integration/Continuous Delivery (CI/CD)} pipelines, where test generation must integrate seamlessly with existing development workflows and maintain consistent quality standards despite the evolving capabilities of underlying LLMs.

In this evolving landscape, previous work provides valuable insights into LLM test generation capabilities, with rigorous evaluations across standardized benchmarks. However, industrial adoption requires addressing additional considerations beyond technical performance metrics. Organizations implementing these tools in production environments must continuously evaluate commercial tools as they evolve, assess compatibility with existing DevSecOps infrastructure \cite{Prates2025}, balance automation benefits against the need for expert oversight, and maintain consistent quality standards across LLM versions. This gap between academic evaluation frameworks and industrial implementation needs motivates our work.

This work presents a practical framework for the continuous evaluation of LLM-based test generators in industrial environments, illustrated through a longitudinal study of GitHub Copilot's test generation capabilities at LKS Next, a medium-sized software consultancy. Our framework integrates with industry-standard tools like SonarQube to provide quantitative quality metrics, implements a continuous re-evaluation pipeline that adapts to evolving LLM capabilities, and addresses practical concerns including test maintainability, cost efficiency, and seamless integration with existing CI/CD processes. This approach helps organizations track the evolution of LLM-based testing tools and assess their practical utility in real development environments.

The key contributions of this work are:
\begin{itemize}
    \item A measurement framework designed specifically for continuous evaluation of commercial LLM test generators.
    \item Integration strategies with industry-standard quality tools and development practices.
    \item Validation through two evaluation cycles showing the evolution of capabilities.
    \item Practical insights for companies seeking to adopt LLM-based testing tools.
\end{itemize}

\section{Background} \label{sec:background}

The evaluation of LLM-generated tests requires understanding the intersection of multiple domains in software engineering. This section examines the foundational concepts of software quality assessment, testing fundamentals, and the emerging role of LLMs in test generation, highlighting how each domain contributes to effective test evaluation.

\subsection{Software Quality Models and Frameworks}
Software quality assessment orchestrates three complementary elements: quality models specify the dimensions and attributes to be evaluated, evaluation frameworks provide the methodological guidance for assessment, and benchmarks establish reference values that give meaning to measurements. For example, when assessing a software system, a quality model (e.g., \cite{iso25010}) identifies key characteristics like maintainability, an evaluation framework prescribes how to measure specific attributes, and benchmarks provide industry standards or historical data for comparison \cite{iso25010, Wagner2012}. 

\subsection{Software Testing Fundamentals}
Software testing is a critical component of quality assurance, requiring both technical expertise and domain knowledge. While metrics like code coverage provide quantitative insights, experienced testers apply sophisticated techniques like equivalence partitioning and boundary value analysis that require human judgment \cite{Fan2019}. Even when automated tools achieve similar coverage metrics to manually written tests \cite{Fan2019}, experts must still evaluate test quality across multiple dimensions including maintainability, readability, and alignment with testing best practices.

\subsection{Large Language Models in Software Engineering}
LLMs have emerged as powerful tools capable of aiding in several software engineering tasks, from code completion to documentation generation \cite{Gao2025, Hou2024}. Consequently, their potential to automatically create tests is being extensively analyzed \cite{Wang24}. However, research shows that automatically generated tests face several challenges:

\begin{itemize}
    \item \textit{Test oracle problem}: Determining correct expected outputs requires domain expertise. \cite{Barr2015}
    \item \textit{Coverage vs. effectiveness gap}: Achieving high coverage does not necessarily guarantee meaningful tests \cite{Kracht2014}.
    \item \textit{Maintainability concerns}: Understanding and evolving generated code requires significant human effort \cite{Fan2019}.
\end{itemize}

These challenges highlight why expert evaluation remains crucial even as automation capabilities advance. Automatically generated tests often contain issues that require human review, including incorrect assertions and problematic test structures \cite{Fan2019}.
\begin{sloppypar}
The intersection of quality models, testing expertise, and LLM capabilities creates both opportunities and challenges for software quality assessment \cite{Gao2025, Sergeyuk2025, Wang24}. While LLMs can accelerate test creation, realizing their benefits in industrial settings requires systematic evaluation frameworks that combine automated metrics with expert assessment of test quality and appropriateness \cite{Kracht2014, Fan2019}.
\end{sloppypar}

\section{Industry Motivation: the Case of LKS Next}

This reseaarch focuses on \textit{LKS Next}\footnote{\url{https://www.lksnext.com}}, a mid-sized technology consultancy that provides software development and technological consulting services to both corporate and public sector organizations. The company specializes in delivering customized software solutions across diverse technologies and programming languages, exemplifying the challenges organizations face when balancing comprehensive testing requirements with market pressures and resource constraints \cite{Garousi2017}.
\begin{sloppypar}
The company's software development lifecycle integrates DevSecOps practices and agile methodologies \cite{Kim2016,Prates2025}, emphasizing automated testing as a cornerstone of their quality assurance strategy. Their continuous integration pipeline implements a systematic validation sequence that includes: compilation and build verification, unit testing of individual components, integration testing for component interactions, automated code quality analysis through SonarQube\footnote{\url{https://www.sonarsource.com/products/sonarqube}}, and dependency scanning for security vulnerabilities \cite{Hilton2016}. This comprehensive automation strategy enables rapid validation of code changes while maintaining consistent quality standards.
\end{sloppypar}
However, like many software development organizations, the company faces increasing pressure to optimize their testing processes while maintaining rigorous quality standards. The resource-intensive nature of test development, particularly for unit and integration tests, presents a significant challenge in their continuous delivery environment \cite{Beller2019,Orso2014}. These challenges make LKS Next a relevant case for studying the adoption of LLM-based testing tools in production environments.

The emergence of AI-powered development tools, particularly LLMs, creates both opportunities and challenges for quality assurance. While these tools can accelerate test generation and reduce developer workload \cite{Wang24}, their effective integration into enterprise environments requires careful consideration of multiple factors. Specifically, the company requires generated tests to adhere to organizational coding standards, architectural patterns, and testing best practices. The company's quality assurance team identified several key requirements, including comprehensive coverage metrics, proper test parameterization, and adherence to testing techniques such as equivalence partitioning and boundary value analysis, along with appropriate use of mocking techniques and test isolation practices for integration testing \cite{Spadini2019}.

Beyond technical considerations, the rapid evolution of LLM capabilities demands continuous evaluation \cite{Gao2025}. As providers frequently release model improvements, quality assurance teams must regularly reassess tool effectiveness, requiring systematic and reproducible evaluation frameworks. Additionally, the pay-per-token model common among LLM providers introduces variable costs that scale with usage. Organizations must balance potential productivity gains against ongoing operational expenses. This balance is particularly critical in continuous integration environments where test execution occurs frequently and at scale \cite{Hilton2016}.

Given these industrial requirements and challenges, we developed a comprehensive evaluation framework with metrics specifically designed to assess LLM-based test generators in production environments. The following section details these metrics and their implementation at LKS Next.

\section{Evaluation Framework: Metrics}

The evaluation of tests generated by LLMs, which we term \textit{AIGen tests (AI-Generated tests)}, requires balancing metrics that can be automatically computed (objective metrics) with those requiring expert judgment (subjective metrics). While objective metrics like compilation success and code coverage can be automatically verified through tools, subjective metrics require expert evaluation to assess aspects like test design quality and methodology alignment with best practices \cite{Fan2019}. This dual approach reflects our earlier observation about the continued importance of expert judgment in test evaluation.

Our framework organizes metrics into three categories:

\begin{itemize}
    \item \textit{Code quality Metrics:} Static analysis of test code structure, from basic compilation to maintainability aspects.
    \item \textit{White Box Testing Metrics:} Runtime coverage and thoroughness measurements.
    \item \textit{Black Box Testing Metrics:} Expert assessment of testing methodology and design. 
\end{itemize}

The first two categories consist primarily of objective metrics that can be automatically measured through standard development tools. The third category requires expert judgment to evaluate testing methodology and design decisions.

\subsection{Code Quality Metrics}

Code quality metrics focus on aspects that can be analyzed without test execution. At the most basic level, test compilation success serves as a gateway - without it, no further evaluation is possible. Beyond compilation, these metrics assess maintainability through static analysis and proper test structure.

Table \ref{tab:compilation-metrics} details the three code quality metrics. First, \textit{Test Compilation Issues} track the accuracy in handling external dependencies and basic syntactic correctness, identifying critical issues that would prevent tests from being integrated into existing codebases. These issues establish the minimum requirements for generated tests to be useful in a production environment. A high number of compilation errors or structural issues indicates that LLM may require additional context or prompt engineering to generate viable tests \cite{Wang24}.

Second, \textit{Static Analysis Issues} and \textit{Setup/Teardown Usage} focus on essential characteristics affecting long-term usability and maintenance of the test suite \cite{Kracht2014, Fan2019}. These aspects are critical in enterprise environments where test maintenance represents a significant portion of testing effort \cite{Garousi2016}. \textit{Static Analysis Issues} are measured using SonarQube's Sonarway profile, enforcing thresholds for maintainability, reliability, and security \cite{campbell2013sonarqube}. \textit{Setup/Teardown Usage} assesses proper test lifecycle management through framework features like @Before and @After annotations, crucial for test independence and reducing duplication \cite{Kim2021}.  High scores in these metrics indicate potential maintenance challenges that could make AIGen tests unsuitable for industrial environments.

\begin{table*}[t]
\caption{Code Quality Metrics}
\label{tab:compilation-metrics}
\begin{tabular}{p{0.20\textwidth} p{0.39\textwidth} p{0.35\textwidth}}
\toprule
\textbf{Metric} & \textbf{Definition} & \textbf{Measurement} \\
\midrule
Test Compilation Issues & Accuracy in handling external library usage and import statements, correct usage of project-specific elements (classes, methods, constructors), basic syntactic and structural correctness of generated test code. & Number of compilation errors in AIGen tests \\
\addlinespace
Static Analysis Issues & Issues identified by SonarQube using the Sonarway profile for the AIGen tests. These issues take into account maintainability, readability, and security in the test code \cite{campbell2013sonarqube}. & Number of static analysis issues found in AIGen tests\\
\addlinespace
Setup/Teardown Usage & Assessment of proper test setup and teardown implementation, focusing on the correct usage of testing framework features (e.g., @Before, @After annotations) for maintainability and efficiency purposes. & (Number of tests with valid setup/teardown by AIGen/ Total number of AIGen tests) × 100 \\

\bottomrule
\end{tabular}
\end{table*}

\subsection{White Box Testing Metrics}

White box testing metrics evaluate how thoroughly the generated tests exercise the code under test. These metrics provide quantitative insights into test effectiveness by measuring different aspects of code coverage, from basic line execution to complex decision paths.

Table \ref{tab:whitebox-metrics} details four metrics that progressively assess test thoroughness. \textit{Line Coverage} provides a basic measure of code exploration, tracking which executable lines are reached during test execution \cite{Swebok2024}. While necessary, line coverage alone is insufficient - tests might execute code without properly verifying its behavior \cite{Kracht2014}.

More sophisticated coverage metrics address this limitation. \textit{Branch Coverage} ensures each boolean sub-expression is evaluated to both true and false, while \textit{Branch/Decision Coverage} combines this with ensuring different execution paths are explored. Together, these metrics help assess whether AIGen tests adequately explore different code paths and boundary conditions.

Finally, \textit{Test Isolation} evaluates whether tests properly manage their dependencies and avoid shared states. This metric is particularly important for integration tests, where improper isolation can lead to flaky tests and maintenance challenges \cite{Spadini2019}. Tests with poor isolation might pass individually but fail when run as part of a larger suite, making them unsuitable for continuous integration environments.

White box metrics are objective and can be automatically computed through standard tools like Jacoco\footnote{https://github.com/jacoco/jacoco}, providing immediate feedback about test thoroughness. High coverage values, while not guaranteeing test effectiveness, indicate that the LLM is capable of generating tests that explore different code paths and conditions. However, these metrics must be considered alongside black box metrics that assess test design quality.

\begin{table*}[t]
\caption{White Box Testing Metrics}
\label{tab:whitebox-metrics}
\begin{tabular}{p{0.20\textwidth} p{0.39\textwidth} p{0.35\textwidth}}
\toprule
\textbf{Metric} & \textbf{Definition} & \textbf{Measurement} \\
\midrule
Line Coverage & Percentage of executable lines of code that have been executed during testing. Ensures basic code path execution \cite{ISO29119}. & (Number of executed lines by AIGen tests / Total executable lines) × 100 \\
\addlinespace
Branch Coverage & Percentage of  branches in the control flow that have been executed during testing \cite{ISO29119}. & (Number of evaluated branches by AIGen tests / Total number of branches) × 100 \\
\addlinespace
Branch/Decision Coverage & Combined measurement ensuring both branch and decisions are thoroughly tested \cite{ISO29119}. & (Number of branches and decisions evaluated by AIGen tests / Total branches and decisions) × 100 \\
\addlinespace
Test Isolation & Assessment of test independence and proper separation of concerns. Measures how well tests avoid shared states and external dependencies \cite{FreemanMPW04}. & (Number of isolated AIGen tests / Total number of tests expected to be isolated) × 100 \\
\bottomrule
\end{tabular}
\end{table*}

\subsection{Black Box Testing Metrics}

Black box testing metrics assess how well AIGen tests implement fundamental testing techniques and align with expert testing practices. Unlike code quality and white box metrics that can be automatically computed, these metrics require expert judgment to evaluate test design decisions and methodology.

Table \ref{tab:blackbox-metrics} details four metrics that evaluate different aspects of test design quality. First, \textit{Equivalence Partitioning Coverage} and \textit{Boundary Value Analysis Coverage} assess the implementation of key testing techniques. Equivalence partitioning evaluates whether tests identify and cover distinct categories of inputs, including both valid and invalid cases \cite{Swebok2024}. Boundary value analysis focuses specifically on edge cases and threshold conditions, measuring if tests exercise these critical scenarios where defects often occur \cite{Swebok2024}.

\textit{Test Parameterization} examines the sophistication of test implementation, assessing whether the LLM recognizes opportunities for parameterization rather than duplicating similar test cases. Well-parameterized tests are more maintainable and provide better documentation of the testing strategy \cite{Tillmann2025}. 

Finally, \textit{Expert-generated Test Coverage} provides a direct comparison against expert-written tests, measuring how well AIGen tests capture the testing patterns and scenarios that experienced developers identify as important. In order to compute this metric, testing experts at LKS Next created manual tests for each analyzed case. This metric helps identify gaps between automated and manual approaches \cite{Kracht2014}, revealing areas where LLM test generation might need enhancement.

These metrics require systematic expert evaluation using predefined criteria to ensure consistent assessment across different LLMs and versions. However, this expert-dependent evaluation is crucial for ensuring that automatically generated tests meet the quality standards required in industrial settings.

\begin{table*}[t]
\caption{Black Box Testing Metrics}
\label{tab:blackbox-metrics}
\begin{tabular}{p{0.19\textwidth} p{0.39\textwidth} p{0.35\textwidth}}
\toprule
\textbf{Metric} & \textbf{Definition} & \textbf{Measurement} \\
\midrule
Equivalence Partitioning Coverage & Assessment of how well tests cover distinct input categories identified through equivalence partitioning. Tests should cover both valid and invalid equivalence classes. & (Number of equivalence partition classes identified by AIGen tests / Total equivalence partition classes identified by experts) \\
\addlinespace
Boundary Value Analysis Coverage & Evaluation of test coverage for boundary conditions, including minimum, maximum, and edge cases. & (Number of boundary values identified by AIGen tests / Total number of boundary values identified by experts) × 100 \\
\addlinespace
Test Parameterization & Measurement of the LLM's ability to generate parameterized tests that efficiently cover multiple test cases. & (Number of parameterized AIGen tests / Number of expected parameterized tests) × 100 \\
\addlinespace
Expert-generated Test Coverage & Percentage of expert-written test scenarios successfully replicated by the LLM, validating its ability to capture human testing knowledge. These tests meet the company's minimum quality standards. & (Number of successfully replicated test scenarios by AIGen tests / Number of test scenarios generated by experts) × 100 \\
\bottomrule
\end{tabular}
\end{table*}

\subsection{Weighted Assessment Framework}

To provide practical guidance on LLM selection and evaluation, our framework combines objective and subjective metrics using a weighted scoring system. This system balances automated measurements with expert assessment to produce a comprehensive quality score, as shown in Table \ref{tab:metric-weights}. It is important to note that the assigned weights reflect LKS Next's specific priorities, and the framework is intentionally designed to be flexible. Other organizations can tailor these weights to align with their own quality standards and strategic goals, while preserving the model's overall structure as a foundation for consistent benchmarking. In our proposal, objective metrics have a weight of 50\% and subjective metrics the other 50\%.

Objective metrics contribute 50\% of the final score, split between code quality and white box metrics:

\begin{itemize}
    \item Code quality metrics can penalize or reward test quality:
        \begin{itemize}
            \item \textit{Compilation Errors (CE, -20\%):} Tests that fail to compile cannot be integrated into the build pipeline, warranting a 20\% penalty. In industrial environments, non-compiling tests block the entire CI/CD process.
            \item \textit{Static Analysis Issues (SAI, -5\%):} Reduces the score by 5\% due to code quality and security violations detected by SonarQube, which indicate potential maintainability, reliability, or security issues. These findings suggest that the current system version is not yet ready for production deployment.
            \item \textit{Setup/Teardown Usage (STU, +10\%):} Rewards proper test structure and maintainability of AIGen test cases in 10\%. 
        \end{itemize}
    \item White box metrics contribute positively (+40\% total). with 10\% allocated to each of the four metrics:
        \begin{itemize}
            \item \textit{Line Coverage (LC, +10\%):} measures executed code lines.
            \item \textit{Branch Coverage (BC, +10\%):} evaluates boolean expression testing.
            \item \textit{Branch/Decision Coverage (BDC, +10\%):} assesses branches and its conditions.
            \item \textit{Test Isolation (TI, +10\%):} measures independence between tests using mocking techniques.
        \end{itemize}
\end{itemize}

Black box metrics constitute the remaining 50\%, reflecting the importance of expert assessment in evaluating testing methodology and test design quality. These include \textit{Equivalence Partitioning Coverage (EPC)}, \textit{Boundary Value Analysis (BVA)}, \textit{Test Parameterization (TP)}, and \textit{Expert-generated Test Coverage (EGTC)}. This equal weighting between objective and subjective metrics ensures that generated tests must meet both technical requirements and testing best practices to achieve a high score.

The final score for each LLM is calculated by combining individual metric scores. For penalty metrics (CE and SAI), we sum the violations across all test cases. For all other metrics, we calculate the average score across examples. Table \ref{tab:metric-weights} provides the detailed formulas for calculating each metric's contribution to the final score.

\begin{table*}[t]
\caption{Metric Weight Distribution and Scoring Formula}
\label{tab:metric-weights}
\renewcommand{\arraystretch}{1.2}
\setlength{\tabcolsep}{4.5pt}  
\begin{tabular}{p{2.8cm}p{3cm}p{1.1cm}p{4.2cm}p{4.4cm}}
\toprule
\textbf{Category} & \textbf{Metric} & \textbf{Weight} & \textbf{Value for LLM} & \textbf{Scoring Formula for LLM} \\
\midrule
\multirow{3}{*}{Code Quality Metrics} 
    & Compilation Errors (CE) & -20\% & $\sum$(CE value for each example) & -(CE value for LLM / MAX(CE value for all LLMs))×20\% \\
    & Static Analysis Issues (SAI) & -5\% & $\sum$(SAI value for each example) & -(SAI value for LLM / MAX(SAI value for all LLMs))×5\% \\
    & Setup/Teardown Usage (STU) & +10\% & $\frac{\sum\text{valid setup/teardown}}{\text{total examples}}$ & (STU value for LLM)×10\% \\
\midrule
\multirow{4}{*}{White Box Metrics} 
    & Line Coverage (LC) & \multirow{4}{*}{+40\%} & $\frac{\sum\text{LC per example}}{\text{total examples}}$ & \multirow{4}{2.5cm}{AVG(LC, BC, BDC, TI)×40\%} \\[2ex]
    & Branch Coverage (BC) & & $\frac{\sum\text{BC per example}}{\text{total examples}}$ & \\[2ex]
    & Branch/Decision Coverage (BDC) & & $\frac{\sum\text{BDC per example}}{\text{total examples}}$ & \\[2ex]
    & Test Isolation (TI) & & $\frac{\sum\text{isolation score per example}}{\text{total examples}}$ & \\
\midrule
\multirow{4}{*}{Black Box Metrics} 
    & Equivalence Partitioning Coverage (EPC) & \multirow{4}{*}{+50\%} & $\frac{\sum\text{EPC per example}}{\text{total examples}}$ & \multirow{4}{2.5cm}{AVG(EPC, BVA, TP, EGTC)×50\%} \\[2ex]
    & Boundary Value Analysis (BVA) & & $\frac{\sum\text{BVA per example}}{\text{total examples}}$ & \\[2ex]
    & Test Parameterization (TP) & & $\frac{\sum\text{TP per example}}{\text{total examples}}$ & \\[2ex]
    & Expert-generated Test Coverage (EGTC) & & $\frac{\sum\text{EGTC per example}}{\text{total examples}}$ & \\
\bottomrule
\end{tabular}
\end{table*}

\section{Evaluation Framework: Methodology}

\begin{figure*}[h]
            \centering
            \includegraphics[width=\textwidth]{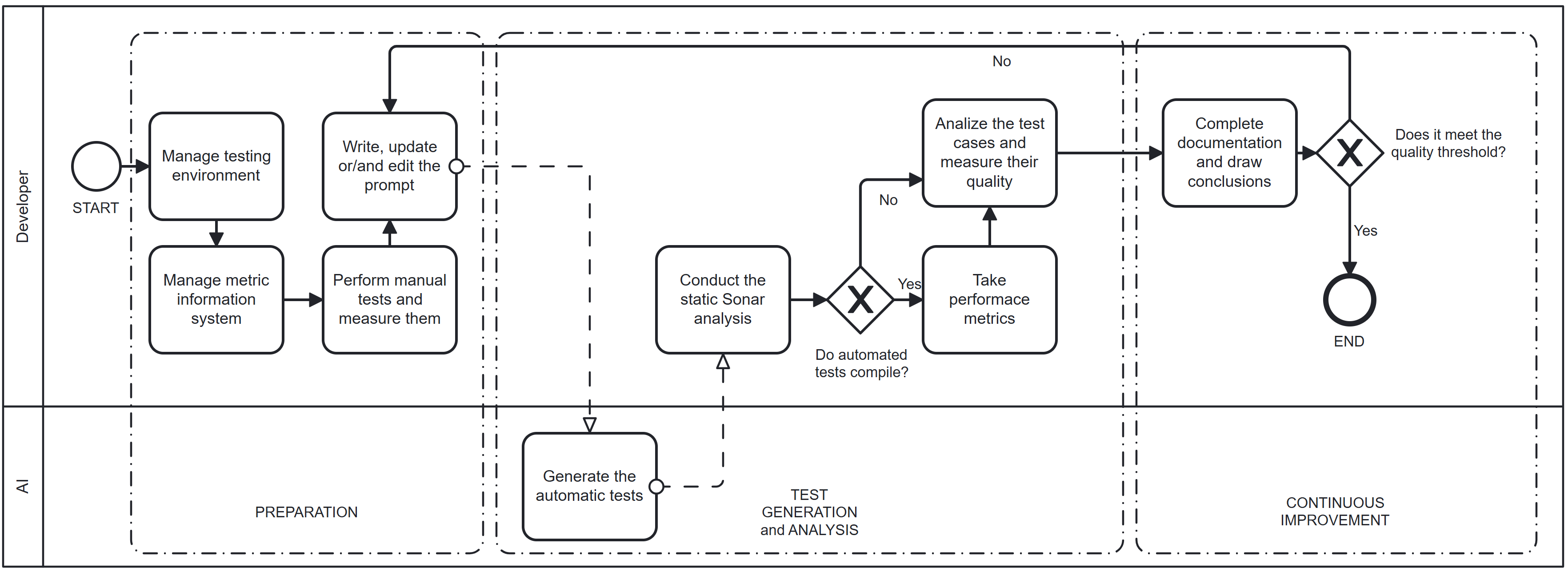}
            \caption{BPMN diagram showcasing the iterative measurement process.}
            \label{fig:bpmn}
\end{figure*}

Our methodology followed an iterative approach for evaluating and improving LLM test generation capabilities. Figure \ref{fig:bpmn} illustrates the complete evaluation workflow, which consists of three main phases: preparation, test generation and analysis, and continuous improvement.

\subsection{Environment Setup and Preparation}

The evaluation process began with environment preparation. This included setting up the IDE, managing project dependencies, and configuring the build tooling. The preparation phase also involved establishing the data collection infrastructure for metric tracking. We used Google Drive for storage with Google Drive Desktop for local replicability, and leveraged Markdown for documentation due to its ease of use and support for code and diagram inclusion.

\subsection{Test Case Selection and Ground Truth}

Selecting appropriate test cases was critical for a meaningful evaluation of the LLM's capabilities in an industrial context. A primary criterion was the avoidance of data leakage. Our initial explorations confirmed previous findings \cite{Lopez2025, Wu2024} that LLMs trained on public repositories often reproduce existing tests rather than generating novel ones, especially when encountering familiar code. To ensure our evaluation measured true generative ability, we deliberately selected functions and classes from projects that did not have pre-existing, publicly accessible test suites.
    
Beyond avoiding data leakage, the core goal was to choose test subjects that represented common challenges and complexities encountered in real-world industrial software development. Important consideration was given to the test compatibility with both unit and integration testing approaches for allowing the evaluation of LLMs across different testing scopes and complexity levels. Due to the significant time requirements of manual expert evaluation needed for measuring subjective metrics, we limited our selection to seven functions. These were specifically chosen to cover a diverse range of typical industrial programming patterns and testing challenges.
    
These functions were chosen not arbitrarily, but specifically for the distinct enterprise-relevant testing scenarios they embodied:

    \begin{enumerate}
        \item To assess the LLM's handling of complex object interactions and the need for mocking external dependencies (like databases), common in multi-layered enterprise applications, we selected the \textit{assemble} method (Rental Assembler). This function requires orchestrating multiple domain objects and simulating database responses.

        \item Representing scenarios involving intricate algorithmic logic and robust exception handling, we included \textit{isPrime} (Prime Number Checker), notable for its high cyclomatic complexity and multiple error paths that demand careful test design.

        \item Core enterprise functionalities often involve user input validation and database interactions. The \textit{addUser} function (User Management) was chosen to test the LLM's ability to handle null conditions, simulate database access (requiring mocking), and manage basic CRUD-like operations.

        \item Business logic frequently involves complex conditional rules and dependencies on other system parts. The \textit{getBonus} function (Bonus Calculator) served as a test case for this, featuring nested conditions and reliance on external data or services.

        \item Enterprise systems often rely on utility functions and require integration testing across multiple small components. The \textit{isIPV4Valid} function (IPV4 Validator) represented this, necessitating tests that validate its interaction with helper functions.

        \item Some enterprise tasks involve specific data structure manipulations. The \textit{isStrobogrammic} function (Strobogrammatic Number) provided a case for testing logic involving map or dictionary structures, using a less widely known algorithm unlikely to suffer from data leakage.

        \item Finally, to evaluate the LLM's capacity to handle integration between multiple internal functions within a class or module, we used \textit{palindrome} (Palindrome Detector). This function's logic depends on other helper methods within the same project, testing the LLM's ability to potentially generate tests for multiple functions in a coordinated manner.
    \end{enumerate}

The full reference to the methods can be found in the replication package\footnote{\url{https://zenodo.org/records/15274212}}. For each of these carefully selected, representative functions, expert Quality Assurance engineers from LKS Next developed comprehensive "ground truth" test suites. These manual tests were crafted following industry best practices and internal quality standards. They were subsequently validated by a QA team leader to ensure full requirements coverage and adherence to code quality metrics, including verification with \textit{SonarQube}. This rigorous process established a reliable benchmark against which the LLM-generated tests could be objectively measured, ensuring our findings reflected performance against realistic, high-quality industrial testing standards.

\subsection{Iterative Evaluation Process}

Following Figure \ref{fig:bpmn}, once the environment and baseline tests were ready, we entered an iterative cycle:

\begin{enumerate}
    \item Generate tests using the current prompt.
    \item Analyze test results through both automated tools and expert review.
    \item Compare results against our quality metrics.
    \item Based on the analysis, either:
        \begin{itemize}
            \item Refine the prompt if improvements are needed.
            \item Document successful results if quality thresholds are met.
        \end{itemize}
\end{enumerate}

This cycle continued until the generated tests met our quality standards or no further improvements could be achieved through prompt refinement.

\subsection{Prompt Engineering Methodology}

Our approach to prompt engineering followed an iterative refinement process based on Prompt Chaining \cite{Wu2022}, which structures generation into sequential components addressing different test creation aspects. Each iteration involved generating tests, evaluating outputs against our quality metrics, identifying weaknesses, and refining prompts to address specific issues.

Our final prompt focused on six key areas identified during evaluation:

\begin{enumerate}
        \item \textit{Addressing Foundational Quality}: To combat compilation errors and static analysis warnings, we introduced explicit instructions demanding adherence to SonarQube rules. The prompt was augmented with directives like: \textit{"When creating the test class, do not generate any line of code that could create any Issue, nor Bug nor Code Smell. I will check with Sonar the code statically..."} Furthermore, we added verification steps requiring the LLM to check imports, constructors, and context before finalizing its response, and crucially, to ask for clarification if context was missing, rather than hallucinating incorrect assumptions.
        
        \item \textit{Mandating Comprehensive Coverage:} Low coverage scores in early iterations prompted the inclusion of specific requirements for testing techniques. We mandated the use of the white box method, focusing on condition/decision coverage, instructing the LLM to "use the truth tables to cover all the cases of the composed condition statements." A non-negotiable target was set: "\textit{I need a 100\% of line, condition and condition/decision coverage}," explicitly mentioning that tools like JaCoCo would be used for verification.
        
        \item \textit{Incorporating Standard Testing Techniques}: Beyond coverage, the quality of test case design was often lacking. To ensure systematic testing, we explicitly required black box testing techniques, namely equivalence partitioning and boundary value analysis, as described above. Recognizing potential overlap between white-box and black-box generated cases, we added a refinement instruction: "If you find redundancy with any of the other methods, remove the redundant tests." For integration tests requiring interaction with dependencies, mocking was mandated using the \textit{Mockito} framework, with specific instructions to avoid direct database interactions to prevent potential ACID errors, instead using mocked classes.
        
        \item \textit{Enforcing Test Engineering Best Practices}: Maintainability and readability issues led to the inclusion of test engineering heuristics. The prompt evolved to include directives such as: "Use proper naming conventions for the tests and structure them correctly... Organize the test class correctly and follow Clean Code rules so the tests are easily understandable and maintainable." To reduce redundancy within the test code itself, the prompt specified the use of \textit{JUnit 5} annotations like \textit{@BeforeAll}, \textit{@BeforeEach}, \textit{@AfterAll}, and \textit{@AfterEach} where appropriate, and the use of \textit{@DisplayName} for clear test descriptions.
    
        \item \textit{Comprehensive Context Provision:} To ensure the LLM could generate both structurally sound and functionally relevant tests, providing comprehensive context was paramount. Our process involved supplying the LLM with the complete source code of the Java class under test (SUT), along with all interconnected classes and interfaces identified through dependency analysis. Crucially, this context included embedded \textit{JavaDoc}, which had been rigorously verified by business domain experts, serving as the official requirements specification. This dual provision of verified requirements (JavaDoc) and full implementation details (source code) enabled the LLM to effectively apply mandated black box techniques based on the specifications, while simultaneously having the necessary visibility to address white box coverage requirements and identifying dependencies for mocking.
    
        \item \textit{Context and Framework Specificity}: To ensure the LLM used the correct libraries and adhered to their specific APIs, the prompt was furnished with the exact Maven dependencies used in the target projects (JUnit 5 Jupiter, Mockito). Links to the official documentation for these frameworks were provided, along with explicit instructions to follow their methodologies. Examples of correct Mockito usage (mock creation, verification, argument captors) were included as templates.
    \end{enumerate}
    
Our analysis revealed significantly better results with English prompts compared to Spanish, aligning with findings from other multilingual studies \cite{Jiang2024, Wang2024exploringmultilingualbiaslarge}. The final optimized prompt (available in our replication package) assigns a specific role to the LLM ("best Software Engineer specialized in testing"), defines tasks, mandates methodologies, enforces quality criteria, and includes self-correction instructions. This structured approach systematically addressed weaknesses identified during evaluation cycles, leading to progressive improvements in test quality as documented in our results.

\subsection{Tooling Infrastructure}

Our evaluation framework leverages industry-standard tools to ensure robust and reproducible assessments:

\begin{itemize}
    \item \textit{Testing Framework}: JUnit for test execution and parameterization.
    \item \textit{Static Analysis}: SonarQube and SonarCloud for code quality metrics.
    \item \textit{Coverage Analysis}: JaCoCo for detailed code coverage measurements.
    \item \textit{Version Control}: GitHub and GitHub Actions for automation.
\end{itemize}

All dependency management was handled through Maven, ensuring reproducible builds and consistent tool versions across evaluations.
    
\section{Results}\label{sec:results}

\begin{table*}[t]
    \centering
    \caption{Metrics Results Comparison}
    \label{tab:testing-metrics}
    \setlength{\tabcolsep}{3pt}  
    \begin{tabular}{l@{\hspace{0.4em}}lccccccc}
        \toprule
            & \textbf{Metric} & \textbf{Weight} & \thead{\textbf{ChatGPT-4}\\\textbf{1st Time}\\(Mar'24)} & \thead{\textbf{ChatGPT-4}\\
\textbf{Iterative}\\(May'24)} & \thead{\textbf{GPT-o}\\(Dec'24)} & \thead{\textbf{o1-}\\
\textbf{Preview}\\(Dec'24)} & \thead{\textbf{o1-}\\
\textbf{Mini}\\(Dec'24)} & \thead{\textbf{Claude 3.5}\\
\textbf{Sonnet}\\(Dec'24)} \\
        \midrule
            \multirow{3}{*}{\textbf{Code Quality Metrics}} & Compilation Errors* & -20\% & 31 & 3 & 2 & 0 & 7 & 0 \\
            & Static Analysis Issues & -5\% & 45 & 18 & 29 & 15 & 10 & 13 \\
            & Setup/Teardown Usage & 10\% & 100.00\% & 100.00\% & 100.00\% & 100.00\% & 85.71\% & 100.00\% \\
        \midrule
            \multirow{4}{*}{\textbf{White Box Testing}} & Line Coverage & \multirow{3}{*}{40\%} & 39.14\% & 65.57\% & 70.14\% & 98.00\% & 28.57\% & 95.71\% \\
            & Branch Coverage & & 39.14\% & 65.57\% & 71.29\% & 95.71\% & 28.57\% & 94.00\% \\
            & Branch/Decision Coverage & & 36.57\% & 61.57\% & 68.29\% & 95.71\% & 28.57\% & 93.29\% \\
            & Test Isolation & & 85.71\% & 100.00\% & 100.00\% & 100.00\% & 100.00\% & 100.00\% \\
        \midrule
            \multirow{4}{*}{\textbf{Black Box Testing}} & Equivalence Partitioning Coverage & \multirow{4}{*}{50\%} & 71.19\% & 75.00\% & 79.88\% & 84.52\% & 85.12\% & 86.90\% \\
            & Boundary Value Analysis Coverage & & 69.39\% & 71.77\% & 78.20\% & 81.53\% & 83.57\% & 83.67\% \\
            & Test Parameterization & & 12.70\% & 38.89\% & 83.81\% & 91.84\% & 88.10\% & 88.89\% \\
            & Expert-generated Test Coverage & & 65.77\% & 75.36\% & 66.23\% & 97.94\% & 81.57\% & 91.48\% \\
        \midrule
            \textbf{Total Weight Assessment} & & & 32.44\% & 67.96\% & 74.97\% & 91.76\% & 63.81\% & 90.72\% \\
        \bottomrule
    \end{tabular}
    \label{tab:results}
\end{table*}

Our longitudinal analysis of LLM-based test generators spans from March 2024 to December 2024, revealing both the rapid evolution of LLM capabilities and the expansion of available models. We focused specifically on GitHub Copilot because it represented the most widely adopted commercial tool leveraging LLMs for software development at the time of our research, with its seamless IDE integration making it particularly relevant for industrial adoption. Our first iteration in March 2024 assessed Copilot when it exclusively used GPT-4, producing largely unusable tests with significant compilation issues. By May 2024, after applying our iterative prompt refinement process, the same model showed substantial improvement. The latest evaluation in December 2024 expanded to include newer models (OpenAI's GPT-o series and Claude 3.5 Sonnet), demonstrating the rapid advancement in this field.

Table \ref{tab:results} presents the complete comparison across models and time periods. Our results reveal substantial improvement across the three key areas.

\textit{Code Quality Metrics} show the most dramatic improvement in basic viability. In March 2024, GitHub Copilot produced an average of 31 compilation errors per test suite and 45 static analysis issues. By May, these numbers dropped to 3 and 18 respectively. The newest models (o1-Preview and Claude 3.5 Sonnet) achieve near-perfect compilation with minimal static issues. Setup/Teardown usage, initially surprising us with good performance (100\%), maintained this level across most models, with only o1-Mini showing slightly lower performance (85.71\%).

\textit{White Box Testing Metrics} demonstrate substantial advancement. Our first tests achieved only 39.14\% line coverage and similar branch coverage levels. The iterative prompt improvement raised these metrics to around 65\%, while newer models reach above 95\% coverage. Particularly noteworthy is the consistent achievement of 100\% test isolation across all recent models, suggesting mature handling of test independence.

\textit{Black Box Testing Metrics} reveal both progress and remaining challenges. While early tests showed reasonable understanding of equivalence partitioning (71.19\%) and boundary analysis (69.39\%), they struggled with test parameterization (12.70\%). The latest models show more balanced performance across all metrics, with significant improvements in test parameterization (reaching 88-91\%) and expert-generated test coverage exceeding 90\% in the best cases.

The total weighted assessment reflects this evolution, with scores rising from 32.44\% in our initial evaluation to over 90\% for the best performing models. This improvement persists across different weighting schemes, suggesting robust advancement in test generation capabilities rather than optimization for specific metrics.

\section{Discussion}

Our evaluation framework has provided valuable insights for both research and industrial practice. The most immediate practical outcome has been its role in the company's decision-making process regarding LLM adoption. In March 2024, when GitHub Copilot generated tests that largely failed to compile and required significant developer intervention to be usable, investing in the technology would have likely increased rather than reduced testing effort. However, the dramatic improvements observed by December 2024 have led the company to reconsider this position, as the latest results suggest potential productivity gains.

Our experience highlights several important considerations for organizations evaluating LLM-based testing tools:

\begin{itemize}
    \item \textit{Continuous Evaluation Required:} The rapid evolution of LLM capabilities means that evaluations quickly become outdated. Organizations need to maintain ongoing evaluation processes rather than relying on point-in-time assessments.
    
    \item \textit{Model Selection and Costs:} While newer models like Claude 3.5 Sonnet and o1-Preview show impressive performance, they come with higher computational costs and latency. Additionally, tools like GitHub Copilot require per-seat licensing, which can be significant for larger teams. Organizations must balance these costs against potential productivity gains.
    
    \item \textit{Technical Investment:} Success with these tools requires investment in two key areas. First, as shown by our results, proper prompt engineering significantly impacts performance. Second, developers need training to effectively use them, particularly in understanding prompt engineering principles and best practices.
    
    \item \textit{Data Privacy Risks:} When testing involves client code or proprietary information, organizations must evaluate how LLMs handle this data. Solutions might include air-gapped environments or local LLM deployments for sensitive code, though these come with their own technical and cost implications.
    
    \item \textit{Expert Oversight:} Despite the high scores achieved in our metrics, expert review remains essential. The black box metrics in particular show that while LLMs have improved in applying testing techniques, they don't yet match human expertise in test design decisions.
\end{itemize}

These findings suggest that while LLM-based test generation tools are becoming increasingly viable for industrial use, their successful adoption requires both systematic evaluation frameworks and realistic expectations. Companies should view these tools as aids to augment developer productivity rather than complete replacements for human testing expertise.

\section{Threats to Validity}
As with any empirical study, particularly one involving emerging technologies like LLMs, there are several potential threats to the validity of our findings. 

\textit{Construct Validity:} Our measurement framework relies on specific metrics to evaluate test quality and LLM effectiveness. There is a risk that these metrics may not fully capture all relevant aspects of test quality or may not accurately reflect the actual value of LLM-generated tests in industrial practice. We mitigate this threat in several ways: (1) combining both automated metrics and expert assessment to provide a more complete evaluation, (2) validating our measurement framework with experienced QA professionals before deployment, and (3) incorporating feedback from practitioners throughout the study period to refine our metrics.

\textit{Internal Validity:} The observed improvements in LLM performance over time could be influenced by factors other than actual evolution in LLM capabilities. In particular, our iterative refinement of prompting strategies might confound the results. To address this, we: (1) maintain consistent base prompts across evaluation periods, (2) clearly document any prompt modifications, and (3) validate significant changes in performance through multiple test runs.

\textit{External Validity:} While our study provides insights from a real-world industrial setting, it is primarily based on experiences from a single organization and specific types of software projects. The effectiveness of our framework and findings may not generalize to organizations with different development practices, project types, or quality requirements. To strengthen external validity, we: (1) designed the framework to be adaptable to different organizational contexts, and (2) focused on industry-standard tools and practices that are common across organizations.

\textit{Reliability:} The expert assessment component of our framework inherently involves subjective judgment, which could affect the reproducibility of our results. We implement two measures to enhance reliability: (1) establishing clear, documented evaluation criteria for expert reviewers, (2) maintaining detailed documentation of all measurement procedures, tools, and decision criteria to enable replication.

While these threats cannot be completely eliminated, we believe our mitigation strategies provide reasonable confidence in our findings.

\section{Related Work}
In the short time since LLMs entered the software development landscape, researchers have explored their potential applications in software testing. Wang et al.'s comprehensive analysis of 102 papers \cite{Wang24} reveals that LLMs are predominantly used for test case preparation (including unit test case generation, test oracle generation, and system test input generation), program debugging, and bug repair. However, their application remains limited in early testing lifecycle activities such as test requirements specification and test planning. The authors identify several key challenges, including: achieving comprehensive test coverage, addressing the test oracle problem, conducting rigorous evaluations, and successfully applying LLMs in real-world testing scenarios. Our work contributes to addressing this last challenge through an industry case study examining the evaluation and adoption of LLM-based testing tools.

Recent empirical studies have begun systematically evaluating LLM capabilities for test generation. Tang et al. \cite{Tang24} present a comparative assessment between ChatGPT and search-based software testing (SBST) tools, finding that while SBST achieves higher code coverage, LLM-generated tests demonstrate promising characteristics in terms of readability and assertion quality. Several research teams have proposed frameworks to enhance LLM-based test generation. Liu et al. introduce AutoTestGPT, which leverages ChatGPT's capabilities through structured prompts and iterative refinement to generate test cases across multiple testing types. Their system demonstrates significant efficiency improvements, reducing framework generation time by over 70\% compared to manual approaches \cite{Liu2024autotestgpt}. Similarly, Chen et al. present ChatUniTest, which addresses key limitations of LLM-based testing through adaptive focal context generation and a generation-validation-repair mechanism. Their evaluation shows ChatUniTest achieving higher line coverage than both traditional and LLM-based alternatives across diverse projects \cite{Chen2024chatunitest}.

Siddiq et al. \cite{Siddiq24} evaluate the unit test generation capabilities of various LLMs (Codex, GPT-3.5-Turbo, and StarCoder) using the HumanEval and EvoSuite SF110 benchmarks. Their results highlight early challenges with coverage, compilability, and test reliability. However, our longitudinal study reveals significant improvements in LLM test generation capabilities over time. Furthermore, we extend beyond prior work by examining integration testing scenarios and evaluating LLMs' ability to generate parameterized tests.

Schäfer et al. \cite{Shafer24} propose an adaptive approach for JavaScript test generation using LLMs. Their TESTPILOT system demonstrates how careful prompt engineering and iterative refinement can improve the quality of generated tests. This aligns with our findings regarding the importance of systematic evaluation frameworks for assessing and improving LLM-based testing tools.

Our work complements these existing studies by providing:
\begin{itemize}
    \item A practical framework for continuous evaluation of LLM test generators in industrial settings.
    \item Insights into the temporal evolution of LLM testing capabilities.
    \item Guidance for companies seeking to adopt LLM-based testing tools while maintaining quality standards.
\end{itemize}

\section{Conclusions}

Our longitudinal evaluation of LLM-based test generators reveals a technology rapidly maturing toward industrial viability. From April 2024 to January 2025, we observed dramatic improvements across all metric categories, particularly in basic test viability and sophistication of testing approaches.

Three key conclusions emerge from our analysis:

\begin{itemize}
    \item \textit{Continuous Evaluation is Essential:} The rapid pace of LLM advancement means that evaluations have a limited shelf life. Our study showed how capabilities can transform within months, from tests that barely compiled to sophisticated test suites achieving over 90\% in our weighted assessment. Organizations need flexible, ongoing evaluation processes to keep pace with these changes.
    
    \item \textit{Human-AI Collaboration:} While newer models show remarkable capabilities, our results emphasize that human oversight remains crucial. The improvements in both objective and subjective metrics suggest LLMs are becoming valuable tools for augmenting, rather than replacing, human expertise in test generation.
    
    \item \textit{Growing Technical Sophistication:} The substantial improvements in test parameterization (from 12.70\% to 88.89\%) and coverage metrics (reaching over 95\%) demonstrate that LLMs can now generate tests that align with industry practices. However, the gap in manual test coverage indicates areas still needing refinement.
\end{itemize}

Several promising directions for future work emerge from these findings. We see value in extending the framework to more complex scenarios, particularly integration and system testing, and investigating how domain-specific prompting affects test quality. Developing automated validation techniques for LLM-generated tests would improve efficiency, while analyzing cost-effectiveness in production environments would provide practical guidance for adoption. Another potential path is exploring hybrid approaches combining LLM capabilities with traditional testing methodologies. Last but not least, analyzing the use of in-house deployed LLMs represents a critical area for investigation, as it could address privacy concerns by preventing sensitive client code or organizational know-how from being shared with third parties.

As LLM capabilities continue to evolve, maintaining updated evaluation frameworks and benchmarks will be crucial. Our results suggest that organizations looking to optimize their testing practices should establish systematic processes for regularly reassessing and integrating these rapidly advancing technologies.

\begin{acks}
This research is co-supported by contract MCIN/AEI/10.13039/50110 0011033 and the European Union NextGeneration EU/PRTR under contract PID2021-125438OB-I00. Funding was also received from the University of the Basque Country (UPV/EHU) under the "University-Enterprise-Society" program (US24/10). Additional support is provided by contract GIU21/037 under the program "Granting Aid to Research Groups at the UPV/EHU (2021)".
\end{acks}

\bibliographystyle{sbc}
\bibliography{2025CIbSE_Bibliography}

\end{document}